\documentclass[titlepage,reprint,aps,pra,showpacs]{revtex4-1}
\usepackage{amsmath,graphicx,dcolumn,bm,xcolor}
\begin{document}
\title{Collisions of three-dimensional bipolar optical solitons in an array of
  carbon nanotubes}
\author{Alexander V. Zhukov and Roland Bouffanais}
\email[Corresponding Author: ]{bouffanais@sutd.edu.sg}
\affiliation{Singapore University of Technology and Design, 8 Somapah Road,
  487372 Singapore}
\author{Boris A. Malomed} \affiliation{Department of Physical Electronics,
  School of Electrical Engineering, Faculty of Engineering, Tel Aviv
  University, 69978 Tel Aviv, Israel}
\author{Herv{\'e} Leblond} \affiliation{LUNAM Universit{\'e}, Universit{\'e}
  d'Angers, Laboratoire de Photonique d'Angers, EA 4464, 2 Boulevard
  Lavoisier, 49000 Angers, France}
\author{Dumitru Mihalache} \affiliation{Academy of Romanian Scientists, 54
  Splaiul Independentei, Bucharest, RO-050094, Romania} \affiliation{Horia
  Hulubei National Institute of Physics and Nuclear Engineering, Magurele,
  RO-077125, Romania}
\author{Eduard G. Fedorov} \affiliation{Department of Biology, Technion-Israel
  Institute of Technology, Haifa 32000, Israel} \affiliation{Vavilov State
  Optical Institute, 199053 Saint Petersburg, Russia}
\author{Nikolay N. Rosanov} \affiliation{Vavilov State Optical Institute,
  199053 Saint Petersburg, Russia} \affiliation{Saint Petersburg National
  Research University of Information Technologies, Mechanics and Optics (ITMO
  University), 197101 Saint Petersburg, Russia} \affiliation{Ioffe
  Physical-Technical Institute, Russian Academy of Sciences, 194021 Saint
  Petersburg, Russia}
\author{Mikhail B. Belonenko} \affiliation{Laboratory of Nanotechnology,
  Volgograd Institute of Business, 400048 Volgograd, Russia}
\affiliation{Volgograd State University, 400062 Volgograd, Russia}
\date{\today }
\begin{abstract}
  We study interactions of extremely short three-dimensional bipolar
  electromagnetic pulses propagating towards each other in an array of
  semiconductor carbon nanotubes, along any direction perpendicular to their
  axes. The analysis provides a full account of the effects of the
  nonuniformity of the pulses' fields along the axes. The evolution of the
  electromagnetic field and charge density in the sample is derived from the
  Maxwell's equations and the continuity equation, respectively. In
  particular, we focus on indirect interaction of the pulses via the action of
  their fields on the electronic subsystem of the nanotube array. Changes in
  the shape of pulses in the course of their propagation and interaction are
  analyzed by calculating and visualizing the distribution of the electric
  field in the system. The numerical analysis reveals a possibility of stable
  post-collision propagation of pulses over distances much greater than their
  sizes.
\end{abstract}
\eprint{PHYSICAL REVIEW A \textbf{94}, 053823 (2016)}
\maketitle
%
%
\section{Introduction}
%

%
Among materials which have been drawing a permanently strong interest in the
course of the last three decades are carbon nanotubes (CNTs). They are
considered as one of the most promising semiconductor materials for developing
a new element base of electronics. CNTs are macromolecular objects, in the
form of layers of graphene rolled into a cylinder, the surface of which is
formed by six carbon cycles~\cite{1}. Nanotubes can be single walled or
multi-walled, if the tube is composed of one or several layers of graphene,
respectively. The CNTs are categorized as chiral, if angles between sides of
the hexagons and nanotube axis are different from 0$^\circ$ and 90$^\circ$, and
achiral otherwise. Further, achiral CNTs exhibit two different structures:
\textquotedblleft saddles," if the sides of the hexagons are perpendicular to
the nanotube axis, or \textquotedblleft zigzags," with a parallel arrangement
of the hexagons relative to the axis.

Since the discovery of nanotubes by Iijima~\cite{2,3} and up to now, a
great deal of work has been done on the synthesis and characterization of
different types of CNTs (see reviews~\cite{4,5,6,7,8} and references
therein). There are detailed descriptions of physical properties of CNTs, as
defined by their geometry and surface structure, i.e., the arrangement of the
hexagonal carbon cycles relative to the axis of the nanotube.  Calculations of
the band structure show that, depending on their build, the CNT may feature
metallic, insulating, or semiconductor properties, that offer a great
potential for applications~\cite{1,4}. Peculiarities of the electron energy
spectrum of semiconductor single-walled CNTs of the zigzag type (see
Refs.~\cite{8,9}) are manifestations of a number of nonlinear electrodynamic
properties similar to those in semiconductors with a superstructure, for
instance, in quantum semiconductor superlattices \cite%
{10,11}. Non-quadratic electron dispersion suggests a possibility of
realization (in electric fields of moderate strength $\sim 10^{3}$--$10^{5}$
V/cm) of diverse nonlinear phenomena, such as nonlinear and absolute negative
conductivity~\cite{12,13}, phase transitions of the first kind induced by the
applied external field~\cite{13}, nonlinear diffraction and self-focusing of
laser beams~\cite{14,15}, electromagnetic solitary waves~%
\cite{16}, etc.

A wide range of applications to modern optoelectronics may stem from these
phenomena. Moreover, the recent advancement of laser physics in the generation
of powerful electromagnetic radiation, including pulses of ultra-short
duration with predetermined parameters~\cite{17,R1,R2,R3,R4,R5}, is an
incentive for a comprehensive study of the propagation of nonlinear
electromagnetic waves, including extremely short pulses, in settings based on
CNTs. It was first predicted theoretically in Ref.~\cite{16} that the
propagation of electromagnetic solitary waves in CNT arrays is possible in a
one-dimensional (1D) model. Later, these results have been extended to more
realistic multidimensional models. In particular, detailed studies of the
propagation of two-dimensional (2D) unipolar and bipolar extremely short electromagnetic pulses
in CNT arrays have been carried out (see
Refs.~\cite{18,19,20}). Three-dimensional (3D)
spatiotemporal optical solitons (``light bullets" \cite{LB1,LB2,LB3}) have
been considered too~\cite{21}.

Actual samples may contain various chemical impurities and structural defects,
both intentionally produced ones or resulting from manufacture
imperfections. The chemical impurities are uniformly distributed over a CNT
array, affecting the dynamics of extremely short electromagnetic pulses \cite%
{22,23,24}. In Refs.~\cite{25,26,27}, a detailed analysis has been developed
for the propagation of unipolar solitary waves in a medium with metallic
inclusions, while Ref.~\cite{28} established the selective nature of the
interaction of an extremely short bipolar electromagnetic pulse with a spot
featuring higher concentration of electrons, induced by a local dopant.
Moving towards a more realistic description of the solitary electromagnetic
waves in CNT arrays, one should gradually increase the complexity of the
underlying model by considering factors that may occur in experimental
situations. Along with the possible presence of static heterogeneities in the
medium, such as, for example, the local inhomogeneity of the
conduction-electron density, it was found necessary to address the shape of
the field of propagating electromagnetic pulses in 2D and 3D models (see
Ref.~\cite{EPJD} and references therein). The distribution of the electric
field along the CNT axis causes, in turn, redistribution of the concentration
of conduction electrons in the medium. Thus, in general, there is another
possible type of heterogeneity of the medium, namely, the dynamic
inhomogeneity induced by the field of the propagating electromagnetic waves.
Effects associated with this type of the induced heterogeneity have never been
thoroughly investigated, to the best of our knowledge.

Obviously, each electromagnetic pulse propagating in the medium is affected by
the spatiotemporal perturbation, induced by the ``trace" of the inhomogeneous
distribution of the electron density caused by the passage of other pulses in
the vicinity of a given one (see Sec.~\ref{sec:electron-density}). Therefore,
it is relevant to consider the propagation of extremely short electromagnetic
pulses in the presence of a dynamical inhomogeneity induced by the fields of
other pulses.  This problem also has relevance to possible applications based
on multiple rapid passages of electromagnetic pulses through the specimen.

In this vein, this work deals with collisions of 3D extremely short
bipolar pulses (light bullets), taking into account the interaction of each
one with perturbations of the electron density induced by the field profile of
the other pulse. The analysis aims to address the collisions in the form close
to that observed in real experiments.

\section{General considerations}


\subsection{Geometry of the problem and restrictions of the model}

\begin{figure}[tbp]
  \includegraphics[width=0.48\textwidth]{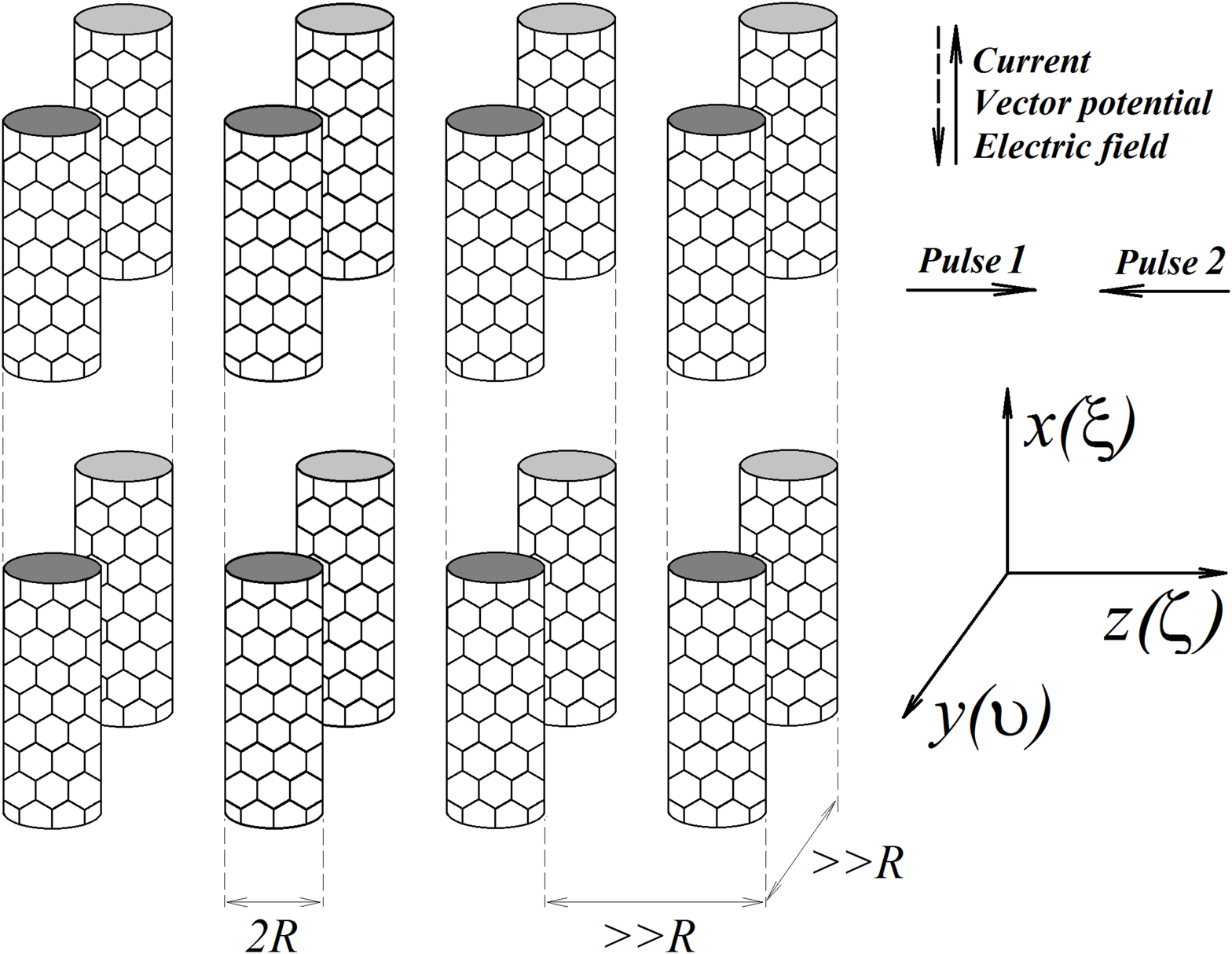}
  \caption{The schematic diagram of the setup with associated coordinate
    system.}
  \label{geometry}
\end{figure}

To begin with, we have to clarify the term \textquotedblleft soliton" used in
this paper. Strictly speaking, we do not provide a mathematical proof for the
ultrashort waves we consider as being solitons (see Appendix A for the outline
of this problem). Specifically, we consider the propagation of a solitary
electromagnetic wave (infrared laser pulse) through a volumetric array of
semiconductor CNTs forming a monolayer of the zigzag type, $(m,0)$, where
integer $m$ (different from a multiple of three) is the number of hexagonal
carbon cycles which form the circumference of the nanotube. The second integer
is the pitch of the helical pattern, also measured as the respective number of
hexagonal carbon cycles, it is zero for a zigzag CNT.  Integer $m$ further
determines the CNT radius, as $R=m\sqrt{3}{b}/{(2\pi )}$%
, where $b=1.42\times 10^{-8}\mathrm{~cm}$ is the distance between
nearest-neighbor carbon atoms~\cite{8,9}. The CNTs are supposed to be placed
into a homogeneous insulator, so that axes of the nanotubes are parallel to
the common $x$ axis, and distances between adjacent nanotubes are much larger
than their diameter, allows one to neglect the interaction between
CNTs~\cite{31} (see Fig.~\ref{geometry}). In particular, with this
configuration one can neglect the electron inter-hopping, supposing that the
corresponding wave functions do not overlap, which is important in order to
avoid uncontrollable transverse currents. The CNT radius is $R\simeq 5.5\times
10^{-8}\mathrm{~cm}$ for $m=7$, which is the value adopted in the computations
below. It is very small in comparison with the radiation wavelength in the
infrared range, which exceeds $10^{-4}\mathrm{~cm}$.

Essentially, the chosen geometry of the problem is similar to the one
considered in Refs.~\cite{21,EPJD}. Given this framework, the dispersion
relation for energy $\epsilon $ of conduction electrons of CNTs is \cite{8,9}
\begin{equation}
  \epsilon (p_{x},s)=\gamma _{0}\sqrt{1+4\cos \left( p_{x}\frac{d_{x}}{\hbar }%
    \right) \cos \left( \pi \frac{s}{m}\right) +4\cos ^{2}\left( \pi \frac{s}{m}%
    \right) },  \label{1}
\end{equation}%
where the electron quasimomentum is $\mathbf{p}=\left\{ p_{x},s\right\} $, $%
s $ being an integer characterizing the momentum quantization along the
perimeter of the nanotube, $s=1,2,\dots ,m$, $\gamma _{0}$ is the overlap
integral, and $d_{x}=3b/2$~\cite{9A}. In this paper, we consider the
propagation of extremely short 3D pulses in direction $z$ perpendicular to
CNT\ axis $x$, with the electric field of the pulses oriented along the $x$
axis (see details below). The duration of each pulse is assumed to be much
smaller than the electron relaxation time $t_{\text{ \textrm{rel}}}$, which
makes it possible to limit the evolution time to $t<t_{\text{\textrm{rel}}}$
[see Eq.  (\ref{t}) below], allowing one to consider the evolution of the
electromagnetic field in the collisionless approximation \cite{21}.

\subsection{Equation for the vector potential}


We treat the electromagnetic field in the CNT array by means of the Maxwell's
equations \cite{32,33} in terms of the vector and scalar potentials,
$\mathbf{A}$ and $\phi $. In the chosen geometry, the governing equation for
the vector potential in the Lorentz gauge is
\begin{equation}
  \frac{\varepsilon }{c^{2}}\frac{\partial ^{2}\mathbf{A}}{\partial t^{2}}-%
  \frac{\partial ^{2}\mathbf{A}}{\partial x^{2}}-\frac{\partial ^{2}\mathbf{A}%
  }{\partial y^{2}}-\frac{\partial ^{2}\mathbf{A}}{\partial z^{2}}=\frac{4\pi
  }{c}\mathbf{j},  \label{2}
\end{equation}%
with $\mathbf{A}=\left\{ A(x,y,z,t),0,0\right\} $, $\mathbf{j}=\left\{
  j(x,y,z,t),0,0\right\} $ is the current density, $\varepsilon $ is the
average relative dielectric constant of the medium \cite{10}, and $c$ is the
speed of light in vacuum. The choice of the vector potential in the CNT
collinear to the axes of the nanotubes is justified by the following
considerations. According to the formulation of the problem, the array has a
nonzero electrical conductivity only along the $x$ axis, while the conduction
current is negligible in the $\left( y,z\right) $ plane, given the vanishingly
small interactions between the nanotubes, hence, the current density is defined
as $\mathbf{j}=\left\{ j(x,y,z,t),0,0\right\} $. In this case, Eq.~\eqref{2}
allows to nullify the second and third components of the vector potential,
$\mathbf{A}=\left\{ A(x,y,z,t),0,0\right\} $.

Expanding the electron energy spectrum \eqref{1} into a Fourier series, and
bearing in mind that the electrons obey the Fermi-Dirac statistics, we apply
the technique developed in Refs.~\cite{34,35}, which makes it possible to
produce an expression for the projection of the current density onto the CNT
axis in the following form (see Appendix B for details):
\begin{equation}
  j=-en\frac{d_{x}}{\hbar }\gamma _{0}\sum_{r=1}^{\infty }G_{r}\sin \left[ r%
    \frac{d_{x}}{\hbar }\left( A\frac{e}{c}+e\int\limits_{0}^{t}\frac{\partial
        \phi }{\partial x}dt^{\prime }\right) \right] ,  \label{3}
\end{equation}%
where $e$ is the electron charge ($e<0$), $n$ the concentration of conduction
electrons in the array, $\phi $ the scalar potential, and coefficients $%
G_{r} $ are given by
\begin{widetext}
\begin{equation}
  G_{r}=-r\frac{\displaystyle\sum\limits_{s=1}^{m}{\displaystyle\frac{\delta _{r,s}}{\gamma _{0}}}\displaystyle%
    \int_{-\pi }^{+\pi }\cos (r\kappa )\left\{ 1+\exp \left[ \frac{\theta
          _{0,s}}{2}+\sum_{q=1}^{\infty }\theta _{q,s}\cos \left( q\kappa \right) \right]
    \right\} ^{-1}\text{d}\kappa }{\displaystyle\sum_{s=1}^{m}\int_{-\pi }^{+\pi
    }\left\{ 1+\exp \left[ \frac{\theta _{0,s}}{2}+\sum_{q=1}^{\infty }\theta _{q,s}\cos
        \left( q\kappa \right) \right] \right\} ^{-1}\text{d}\kappa }.  \label{4}
\end{equation}%
\end{widetext}
Here, $\theta _{r,s}=\delta _{r,s}(k_{B}T)^{-1}$, $T$ is the temperature, $%
k_{B}$ the Boltzmann constant, and $\delta _{r,s}$ are coefficients of the
Fourier decomposition of spectrum~\eqref{1}:
\begin{equation}
  \delta _{r,s}=\frac{d_{x}}{\pi \hbar }\int_{-\pi \hbar /d_{x}}^{-\pi \hbar
    /d_{x}}\epsilon (p_{x},s)\cos \left( r\frac{d_{x}}{\hbar }p_{x}\right) \text{
    d}p_{x}.  \label{5}
\end{equation}%
Next, we combine Eqs.~\eqref{2} and~\eqref{3} to derive an equation governing
the evolution of the vector potential in the CNT array, in the dimensionless
notation:
\begin{widetext}
\begin{equation}
  \frac{\partial ^{2}\Psi }{\partial \tau ^{2}}-\left( \frac{\partial ^{2}\Psi
    }{\partial \xi ^{2}}+\frac{\partial ^{2}\Psi }{\partial \upsilon ^{2}}+\frac{%
      \partial ^{2}\Psi }{\partial \zeta ^{2}}\right) +\eta \sum_{r=1}^{\infty
  }G_{r}\sin \left[ r\left( \Psi +\int\limits_{0}^{\tau }\frac{\partial \Phi }{%
        \partial \xi }d\tau ^{\prime }\right) \right] =0,  \label{6}
\end{equation}%
\end{widetext}
where $\eta =n/n_{0}$ is the scaled electron concentration, $n_{0}$ being the
equilibrium concentration in a homogeneous specimen in the absence of the
electromagnetic field, $\Psi =Aed_{x}/\left( c\hbar \right) $ is the
projection of the scaled vector potential onto the $x$ axis, $\Phi =\phi
\sqrt{\varepsilon }ed_{x}/(c\hbar )$ is the dimensionless scalar potential, $%
\tau =\omega _{0}t/\sqrt{\varepsilon }$ is the scaled time, $\xi =x\omega
_{0}/c$, $\upsilon =y\omega _{0}/c$ and $\zeta =z\omega _{0}/c$ are the scaled
coordinates, and
\begin{equation}
  \omega _{0}\equiv 2\frac{|e|d_{x}}{\hbar }\sqrt{\pi \gamma _{0}n_{0}}.
  \label{7}
\end{equation}%
Equation (\ref{7}) determines a characteristic angular frequency of the CNT
electron subsystem in the conduction band, which is similar to the plasma
frequency of electrons in semiconductor superlattices, cf. Ref.~\cite{10}.

\subsection{The equation for the scalar potential field}


Fields of extremely short electromagnetic pulses under consideration are
localized in all the three directions of the Cartesian coordinates system.
The nonuniformity of the field along the direction perpendicular to the CNT
axis has no impact on the distribution of the electron concentration in the
sample, as the interaction between the nanotubes is negligible, and, as said
above, there is no electric current in the $\left( y,z\right) $ plane. On the
contrary, the field nonuniformity along the $x$ axis perturbs the
conduction-current density, which, in turn, affects the charge density in the
sample~\cite{21,EPJD}. In this connection, we note that concentration $n$ of
the conduction electrons in the expression for the conduction current (%
\ref{3}) and, consequently, the scaled electron concentration, $\eta $, are,
in general, functions of the coordinates and time, $\eta =\eta (\xi ,\upsilon
,\zeta ,\tau )$.

Obviously, the redistribution of the electron density leads to a change of the
scalar potential. The Maxwell's equations \cite{32,33} produce an evolution
equation for the scalar potential (see also Ref.~\cite{21}):
\begin{equation}
  \frac{\partial ^{2}\Phi }{\partial \tau ^{2}}-\left( \frac{\partial ^{2}\Phi
    }{\partial \xi ^{2}}+\frac{\partial ^{2}\Phi }{\partial \upsilon ^{2}}+\frac{
      \partial ^{2}\Phi }{\partial \zeta ^{2}}\right) =\beta (\eta -1),  \label{8}
\end{equation}%
where $\beta =c\hbar /\left( d_{x}\gamma _{0}\sqrt{\varepsilon }\right) $.

\subsection{Equation for the electron density}
\label{sec:electron-density}

As mentioned above, the nonuniformity of the electric field along the CNT\
axis ($x$) perturbs the current density in this direction, as follows from
Eq.~\eqref{3}, leading to redistribution of the electron density. The total
charge in the sample being conserved, the change in the bulk charge density,
$\rho =en$, obeys the continuity equation, $\nabla \cdot \mathbf{j}%
+\partial \rho /\partial t=0$ \cite{32,33}. Projected onto the CNT\ axis, this
equation reads as
\begin{equation}
  \frac{\partial j}{\partial x}+\frac{\partial \rho }{\partial t}=0.  \label{9}
\end{equation}%
Substituting Eq.~\eqref{3} into Eq.~\eqref{9}, and passing to the
dimensionless notation, we obtain an evolution equation for the electron
concentration under the action of the pulse's electromagnetic field:
\begin{equation}
  \frac{\partial \eta }{\partial \tau }=\alpha \sum_{r=1}^{\infty }G_{r}\frac{%
    \partial }{\partial \xi }\left\{ \eta \sin \left[ r\left( \Psi
        +\int\limits_{0}^{\tau }\frac{\partial \Phi }{\partial \xi }d\tau ^{\prime
        }\right) \right] \right\} ,  \label{10}
\end{equation}%
where $\alpha \equiv d_{x}\gamma _{0}\sqrt{\varepsilon }/c\hbar $. It is worth
noting that the nonlinearity of the concentration of electrons [see
Eq.~\eqref{10}] bears the same nature as the nonlinearity of the current
density [see Eq.~\eqref{3}], the latter being responsible for the reshaping of
the electron concentration. Furthermore, the accumulation of charge is a
direct consequence of the inhomogeneity of the current along the
CNTs. Accordingly, an additional electric field can appear, which is fully
accounted for in this study.

Thus, the evolution of the field in the array, taking into regard the
perturbation of the conduction-electron density due to the nonuniformity of
the field along the CNT\ axis, is governed by Eqs. \eqref{6}, \eqref{8}, and %
\eqref{10}. This is a self-consistent system for the coupled evolution of the
field and electron density: the field impacts the dynamics of electrons, and
the latter's feedback affects the evolution of the field, which resembles the
Vlasov's equations in plasma physics \cite{Vlasov}, and the recently studied
local-field effect for the propagation of optical and microwave fields in
atomic Bose-Einstein condensates \cite{Shanghai}.

\subsection{Characteristics of the pulse field}


Measuring instruments can record the energy characteristics of the pulse
defined by the electric field \cite{37}. The electric field in the CNT\ array
is determined by the potentials, $\displaystyle\mathbf{E}%
=-c^{-1}\partial \mathbf{A/\partial t}-\nabla \phi $ \cite{32,33}, which can
be written using the dimensionless variables defined above:%
\begin{equation}
  \mathbf{E}=E_{0}\left( \frac{\partial \Psi }{\partial \tau }+\frac{\partial
      \Phi }{\partial \xi },\frac{\partial \Phi }{\partial \upsilon },\frac{%
      \partial \Phi }{\partial \zeta }\right) ,  \label{11}
\end{equation}%
\begin{equation}
  E_{0}\equiv -\frac{\hbar \omega _{0}}{ed_{x}\sqrt{\varepsilon }}.  \label{12}
\end{equation}%
Thus, Eq.~\eqref{11} demonstrates that the electric field can be represented
as a superposition of two components, $\mathbf{E}=\mathbf{E}_{\Vert }+%
\mathbf{E}_{\perp }$, where $\mathbf{E}_{\Vert }$ is directed along the CNT
axis,
\begin{equation}
  \mathbf{E}_{\Vert }=E_{0} \left( \frac{\partial \Psi }{\partial \tau }+%
    \frac{\partial \Phi }{\partial \xi },0,0\right) ,  \label{13}
\end{equation}%
and $\mathbf{E}_{\perp }$ is the electric field in the orthogonal plane,
\begin{equation}
  \mathbf{E}_{\perp }=E_0\left( 0,\frac{\partial \Phi }{\partial \upsilon
    },\frac{\partial \Phi }{\partial \zeta }\right) .  \label{14}
\end{equation}

Electric field $\mathbf{E}_{\perp }$ has no effect on the dynamics of
electrons due to the absence of the conductivity in the $\left( y,z\right) $
plane. Expression \eqref{14} shows that field $\mathbf{E}_{\perp }$ appears
due to the perturbation of the conduction-electrons density and the presence
of the scalar potential. This field is determined by the projection of vector
$-\nabla \phi $ onto the $(y,z)$ plane. In other words, the nonuniformity of
field $\mathbf{E}_{\Vert }$ along the CNT\ axis gives rise to electric field
$\mathbf{E}_{\perp }$ in the orthogonal plane.

As the conductivity of the array is different from zero only along the $x$
axis, the dynamics of the electron subsystem affects only electric field $%
\mathbf{E}_{\Vert }$, which, in turn, is itself generated by the dynamics of
electrons. The interaction of field $\mathbf{E}_{\Vert }$ with the electron
subsystem gives rise to a self-consistent field of the electromagnetic
solitary wave. To visualize the distribution of the wave field and identify
its localization, we introduce $I=I(\xi ,\upsilon ,\zeta ,\tau )\equiv |%
\mathbf{E}_{\Vert }|^{2}/\left( 8\pi \right) $. It has the dimension of the
volume energy density \cite{32,33}, therefore we call it the bulk energy
density of the electric field of the wave. Following Eqs.~\eqref{12} and %
\eqref{13}, it may be expressed as
\begin{equation}
  I=I_{0}\left( \frac{\partial \Psi }{\partial \tau }+\frac{\partial \Phi }{%
      \partial \xi }\right) ^{2},  \label{15}
\end{equation}%
where $I_{0}=(\hbar \omega _{0})^{2}/(e^{2}d_{x}^{2}8\pi \varepsilon )$.
Numerical calculations reveal that the profile of $I(\xi ,\upsilon ,\zeta
,\tau )$ may feature pronounced maxima, whose positions at any given moment in
time are identified as position of the solitary waves.

\subsection{Methodological aspects}


Concluding this section, we would like to specify two aspects which, from a
methodological point of view, may be useful for modeling wave phenomena in
nonlinear media. In general, there are at least two different approaches to
the analysis of the problem. The primary focus here is on the mathematical
scheme based upon the Fourier expansion of dispersion relation~\eqref{1} (see
Appendix B), and we here explain the feasibility of this approach.  Current
density~\eqref{3} determines Eq.~\eqref{6} for the vector potential
field, and Eq.~\eqref{10} for the electron concentration. We stress that
the expression for current density~\eqref{3}, and therefore Eqs. %
\eqref{6} and \eqref{10}, does not contain any explicit dependencies following
from the particular form of function $\epsilon (p_{x},s)$. In other words, for
any medium, defined by its electron dispersion, which may be different from
Eq.~\eqref{1}, the current density will be expressed similarly to Eq.~%
\eqref{3}. At the same time, properties of the specific environment,
determined by its electron energy spectrum, will affect the expression for the
current density through coefficients similar to $G_{r,s}$. Thus, deriving
equations describing the evolution of electromagnetic waves in an the CNT\
array, we also derive a universal approach for optimizing simulations of the
wave propagation in generic media defined by the energy dependence of the
conduction electrons on their quasimomenta. Given that the equations for the
current density and vector potential have been derived, one can select a
numerical scheme for solving the corresponding system of equations, this
scheme remaining effective for other environments. Of course, a particular
medium comes with a specific electron-energy spectrum, thus leading to
different values of coefficients similar to those in Eq.~%
\eqref{4}, which determine the Fourier decomposition of the spectrum.
Nonetheless, the actual form of the governing equation [similar to Eqs.~%
\eqref{6} and~\eqref{10}], and the type of the numerical scheme remain the
same.

There is another approach to the visualization of the pulse evolution, which
can be combined with the one proposed above. Intensity $I(\xi ,\upsilon ,\zeta
,\tau )$ given by Eq.~\eqref{15} yields the distribution of the electric-field
energy of the pulse at a fixed instant of time. However, neither the
distribution of the electric field, nor the distribution of the field's energy
density at a given moment of time determine the direction of motion of
electromagnetic pulses. In other words, when plotting the distribution $I(\xi
,\upsilon ,\zeta ,\tau )$ at a particular moment of time, one cannot directly
know in what direction the electromagnetic waves propagate, without having at
least one similar plot at another moment of time. However, there is a method
for identifying the propagation direction of the electromagnetic pulse at time
$\tau $ without calculating the distribution of $I(\xi ,\upsilon ,\zeta ,\tau
)$ at time $\tau +d\tau $.  Namely, one can supplement $I(\xi ,\upsilon ,\zeta
,\tau )$ at time $\tau $ by Poynting vector
$\mathbf{S}=\{S_{x},S_{y},S_{z}\}$~\cite{32,33} (see Appendix C).

The absolute value of the Poynting vector determines the magnitude of power
flux carried by the field at each point of the CNT\ array at any instant of
time. The direction of the power transfer is determined by the sign and
magnitude of components of $\mathbf{S}$; $\text{sign}(S_{x})$, $\text{sign}%
(S_{y})$, $\text{sign}(S_{z})$ and $|S_{x}|$, $|S_{y}|$, $|S_{z}|$,
respectively. To explain this possibility, we resort to the simple example of
the wave propagating along the $z$ axis. The propagation of the
electromagnetic pulse in the positive direction of $z$ is associated with the
transfer of energy in this direction, which implies a positive Poynting-vector
component $S_{z}$. Accordingly, the propagation in the negative direction is
associated with $S_{z}<0$. Thus, quantities $I(\xi ,\upsilon ,\zeta ,\tau )$
and $S_{z}(\xi ,\upsilon ,\zeta ,\tau )$ complement each other in drawing the
complete structure of the electromagnetic field at any given moment: $I(\xi
,\upsilon ,\zeta ,\tau )$ determines the spatial localization of the
electromagnetic field (i.e., the state of the system at time $\tau $), while a
particular component of the Poynting vector identifies the direction and
intensity of the field-energy transfer along the respective axis, making it
possible to predict the state of the system at time $\tau +d\tau $.

\section{Numerical results}


\subsection{System's parameters and initial conditions}


We assume that initially (at time $\tau =\tau _{0}$) the electron density is
uniform with value $n_{0}$, while the scalar potential is zero throughout the
sample. These initial conditions are similar to those used in Ref.~\cite%
{21}:
\begin{align}
  \eta (\xi ,\upsilon ,\zeta ,\tau _{0})& =1,  \label{16} \\
  \Phi (\xi ,\upsilon ,\zeta ,\tau _{0})& =0.  \label{17}
\end{align}%
Assuming that there are two electromagnetic pulses propagating towards each
other in the CNT array, we define the initial projection of the vector-
potential field of the pair of pulses as follows:
\begin{align}
  \Psi (\xi ,\upsilon ,\zeta ,\tau _{0})=\nonumber\\\sum_{i=1}^{2}\left\{ \Psi _{i}(\zeta
    ,\tau _{0})\exp \left[ -\frac{(\xi -\xi _{0i})^{2}+(\upsilon -\upsilon
        _{0i})^{2}}{\lambda _{i}^{2}}\right] \right\} ,  \label{18}
\end{align}%
where $\Psi _{i}(\zeta ,\tau _{0})$ is the corresponding profile of the
projection of the $i$-th pulse onto the $\xi $ axis at $\xi =\xi _{0i}$ and $%
\upsilon =\upsilon _{0i}$, and $\lambda _{i}$ is the dimensionless initial
transverse half-width of the $i$-th pulse, while $\xi _{0i}$ and $\upsilon
_{0i}$ are the coordinates of the pulses' centers at $\tau =\tau _{0}$. We
have chosen the Gaussian profile of the input, given by Eq.~\eqref{18}, in the
$\left( \xi ,\upsilon \right) $ plane due to its occurrence in various
applications~\cite{37,38,14}.

To further justify the choice of the initial profile of each electromagnetic
pulse in the longitudinal ($\zeta $) direction, we provide the following
arguments relating to Eq. \eqref{6}. First, as shown by numerical
calculations, coefficients $G_{r}$ [see Eq.~\eqref{4}] rapidly decay with the
increase of $r$. Therefore, we keep only the terms with $r=1$ in Eq.~%
\eqref{6}. Second, for the time being we only consider the variation of the
field in the longitudinal direction, restricting ourselves to the 1D
description with coordinate $\zeta $ and accepting the assumption of
uniformity of the field along the $\xi $ and $\upsilon $ axes. Third, the
assumption of the uniformity of the field along the $\xi $ axis allows us to
neglect possible perturbation of the electron density (i.e., we have $\eta
\approx 1$). It follows from here that a single scalar potential keeps a
constant value throughout the sample, i.e., $\partial \Phi /\partial \xi =0$%
. This conclusion also follows from Eq.~\eqref{8}, which for $\eta =1$ admits
the trivial stationary solution, $\Phi =0$. As a result, we reduce
Eq.~\eqref{6} to an evolution equation for the nonvanishing component of the
vector potential:
\begin{equation}
  \frac{\partial ^{2}\Psi }{\partial \tau ^{2}}-\frac{\partial ^{2}\Psi }{%
    \partial \zeta ^{2}}+\sigma ^{2}\sin \Psi =0,  \label{19}
\end{equation}%
where we define $\sigma \equiv \sqrt{G_{1}}$. Our calculations always produce
$G_{1}>0$, hence, $\sigma $ is real, with $\sigma ^{2}>0$. Equation %
\eqref{19} has the form of the celebrated sine-Gordon equation, which gives
rise to solutions in the form of a breather, i.e., an oscillating
nontopological soliton \cite{39}:
\begin{widetext}
\begin{equation}
  \Psi _{B}(\zeta ,\tau )=4\arctan \left\{ \sqrt{\frac{1}{\Omega ^{2}}-1}\frac{%
      \sin \left( \sigma \Omega \frac{\tau -(\zeta -\zeta _{0})u/v}{\sqrt{%
            1-(u/v)^{2}}}\right) }{\cosh \left( \sigma \left[ \tau u/v-(\zeta -\zeta
          _{0})\right] \sqrt{\frac{1-\Omega ^{2}}{1-(u/v)^{2}}}\right) }\right\} ,
  \label{20}
\end{equation}%
\end{widetext}
with $\Omega \equiv \omega _{B}/\omega _{0}<1$, $u$ being the speed of the
pulse propagation, $v=c/\sqrt{\varepsilon }$ the linear speed of the
electromagnetic waves in the medium, while $\zeta _{0}$ is the coordinate of
the breather's center at $\tau =\tau _{0}$.

Note that Eq.~\eqref{6} shares similarities with Eq.~\eqref{19}, and, in some
sense, it may be treated as a non-1D inhomogeneous modified sine-Gordon
equation. Since Eq.~\eqref{19} generates breather solutions given by Eq. %
\eqref{20}, one may expect a possibility for the propagation of similar
solitary waves generated by Eq.~\eqref{6}.

We assume that the CNT\ array is irradiated by two bipolar ultra-short
electromagnetic pulses propagating towards each other, so that the
vector-potential field of the pair of pulses is determined by Eq.~\eqref{18}%
, in which $\Psi _{i}(\zeta ,\tau _{0})$ have a form similar to that in Eq.~%
\eqref{20}:
\begin{equation}
  \Psi _{i}(\zeta ,\tau _{0})=4\arctan \left\{ \sqrt{\frac{1}{\Omega ^{2}}-1}%
    \frac{\sin \chi _{i}}{\cosh \mu _{i}}\right\} ,  \label{21}
\end{equation}%
\begin{align}
  \chi _{i}& \equiv \sigma \Omega _{i}\frac{\tau _{0}-(\zeta -\zeta
    _{0i})u_{i}/v}{\sqrt{1-(u_{i}/v)^{2}}},  \label{22} \\
  \mu _{i}& \equiv \sigma \left[ \frac{\tau _{0}u_{i}}{v}-(\zeta -\zeta
    _{0i})%
  \right] \sqrt{\frac{1-\Omega _{i}^{2}}{1-(u_{i}/v)^{2}}}.  \label{23}
\end{align}%
Here we use the term \textquotedblleft bipolar" in the sense that the
electromagnetic field changes its sign in the course of the pulse
propagation. We emphasize that we choose the far separated initial (at $\tau
=\tau _{0}$) electromagnetic pulses, given\ by initial conditions~\eqref{21}%
, hence the interaction between them is initially negligible.

It is worth stressing that the use of 1D equation~\eqref{19} relates solely
to\ the choice of the longitudinal profile of the initial
condition [specifically, in the form of Eq.~\eqref{20}] while we stay in
the general framework of the 3D model. Clearly, the system of governing
equations \eqref{6}, \eqref{8}, and \eqref{10}, with initial conditions %
\eqref{16}, \eqref{17}, \eqref{18}, and \eqref{21}, does not have exact
analytical solutions. We have therefore conducted a numerical investigation of
the interaction of the electromagnetic pulses in the CNT array. To simulate
the system of governing equations, we employed an explicit finite-difference
scheme previously used and detailed in Refs.~\cite{EPJD,40}, which was
generalized for this 3D setting. Difference scheme steps in both time and
space were iteratively decreased twice until the solution became unchanged in
the eighth decimal place, thereby ensuring both spatial and temporal
convergence of the obtained solution. Thus, we have numerically found values
of $\Psi (\xi ,\upsilon ,\zeta ,\tau )$, $\Phi (\xi ,\upsilon ,\zeta ,\tau )$,
$\eta (\xi ,\upsilon ,\zeta ,\tau )$, the energy density of the electric
field, defined as per Eq. \eqref{15}, and also the Poynting vector.

In our simulations, we used the following realistic values of the parameters:
$m=7$, $b=1.42\times 10^{-8}$ cm, $\gamma _{0}=2.7$ eV, $%
d_{x}\approx 2.13\times 10^{-8}$ cm. The system is immersed into a medium with
the relative dielectric constant $\varepsilon =4$. We consider the system at
room temperatures $T=293$~K, with the equilibrium electron concentration
$n_{0}=10^{18}$ cm$^{-3}$ \cite{31}. As follows from Eq.~%
\eqref{7}, $\omega _{0}\approx 7.14\times 10^{13}\mathrm{~rad\,s}^{-1}$, $%
\alpha $ and $\beta $ in Eqs.~\eqref{8} and \eqref{10} take values $\alpha
\approx 5.8\times 10^{-3}$ and $\beta \approx 1.72\times 10^{2}$, while $%
\sigma =0.95$ in Eqs.~\eqref{22} and \eqref{23}. Note also that our results
are obtained in the framework of the collisionless model, which is valid at
times not exceeding the above-mentioned relaxation time:
\begin{equation}
  t<t_{\text{\textrm{rel}}}\approx 3\times 10^{-13}\mathrm{~s}  \label{t}
\end{equation}%
\cite{9}, which may be sufficient to complete the collision between the
pulses.

The parameters of the electromagnetic pulses, $u_{i}/v$, $\Omega _{i}$, and $%
\lambda _{i}$, were varied in a wide range, similar to what was done
previously in Ref.~\cite{21}. To be specific, we here present typical results
for the following initial parameters: $u_{1}/v=-u_{2}/v=0.95$, $%
\Omega _{1,2}=0.5$, which corresponds to a half-cycle pulse with vacuum
wavelength $\approx 16\mu \mathrm{m}$, belonging to the long-wavelength
infrared range, and $\lambda _{1,2}=2$, $\xi _{01,2}=\upsilon _{01,2}=0$, $%
\zeta _{02}=-\zeta _{01}=3$. Note that, because of our choice of the initial
conditions, the centers of the electromagnetic pulses, $(\xi _{01},\upsilon
_{01},\zeta _{01})$ and $(\xi _{02},\upsilon _{02},\zeta _{02})$, are
initially located on the $\zeta $ axis.

\subsection{Interactions of the electromagnetic pulses}

\begin{figure}[tbp]
  \includegraphics[width=0.48\textwidth]{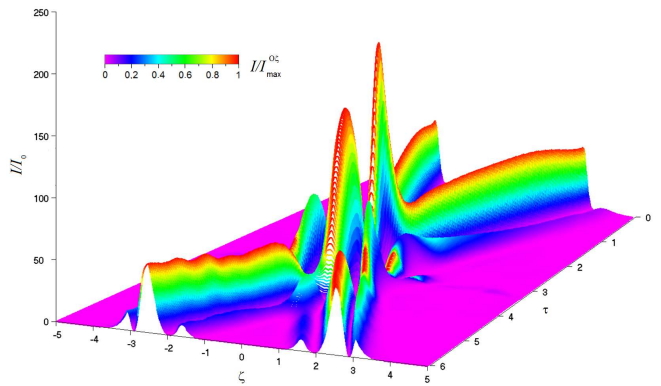}
  \caption{The evolution of the energy density $I(\protect\xi =0,\protect%
    \upsilon =0,\protect\zeta ,\protect\tau )$ of the electric field on the $%
    \protect\zeta $ axis. Values of $I$ at every instant of time are
    color coded, with red and purple corresponding to maxima and minima.}
  \label{fig1}
\end{figure}
\begin{figure}[tbp]
  \includegraphics[width=0.48\textwidth]{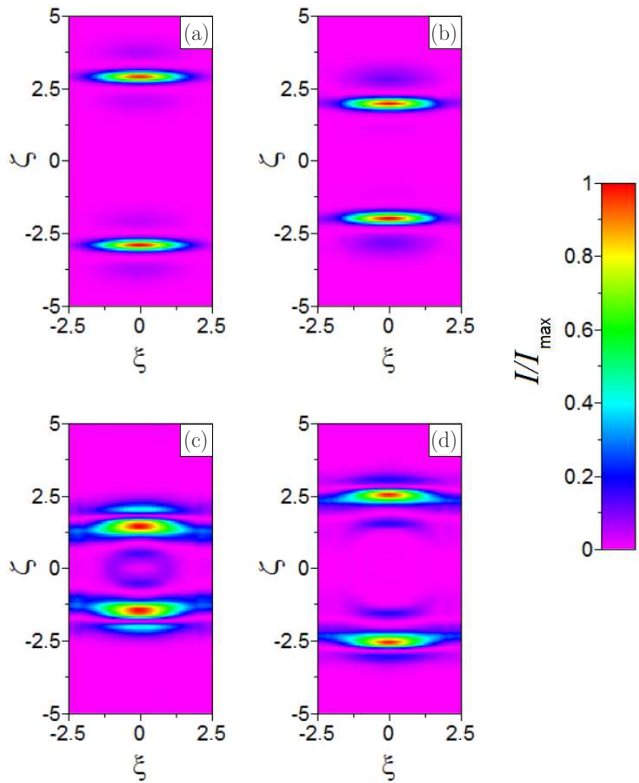}
  \caption{The distribution of the energy density $I(\protect\xi ,0,\protect%
    \zeta ,\protect\tau )$ of the electric field in the plane of $\left(
      \protect%
      \xi ,\protect\zeta \right) $ prior to and past the collision, at
    different times $\protect\tau$: (a) $\protect\tau =0.1$, (b)
    $\protect\tau =1.0$, (c) $\protect\tau =5.0$, (d) $\protect\tau
    =6.0$. Values of $I$ are color coded in the same way as in
    Fig.~\protect\ref{fig1}.}
  \label{fig2}
\end{figure}
\begin{figure*}[tbp]
  \includegraphics[width=0.9\textwidth]{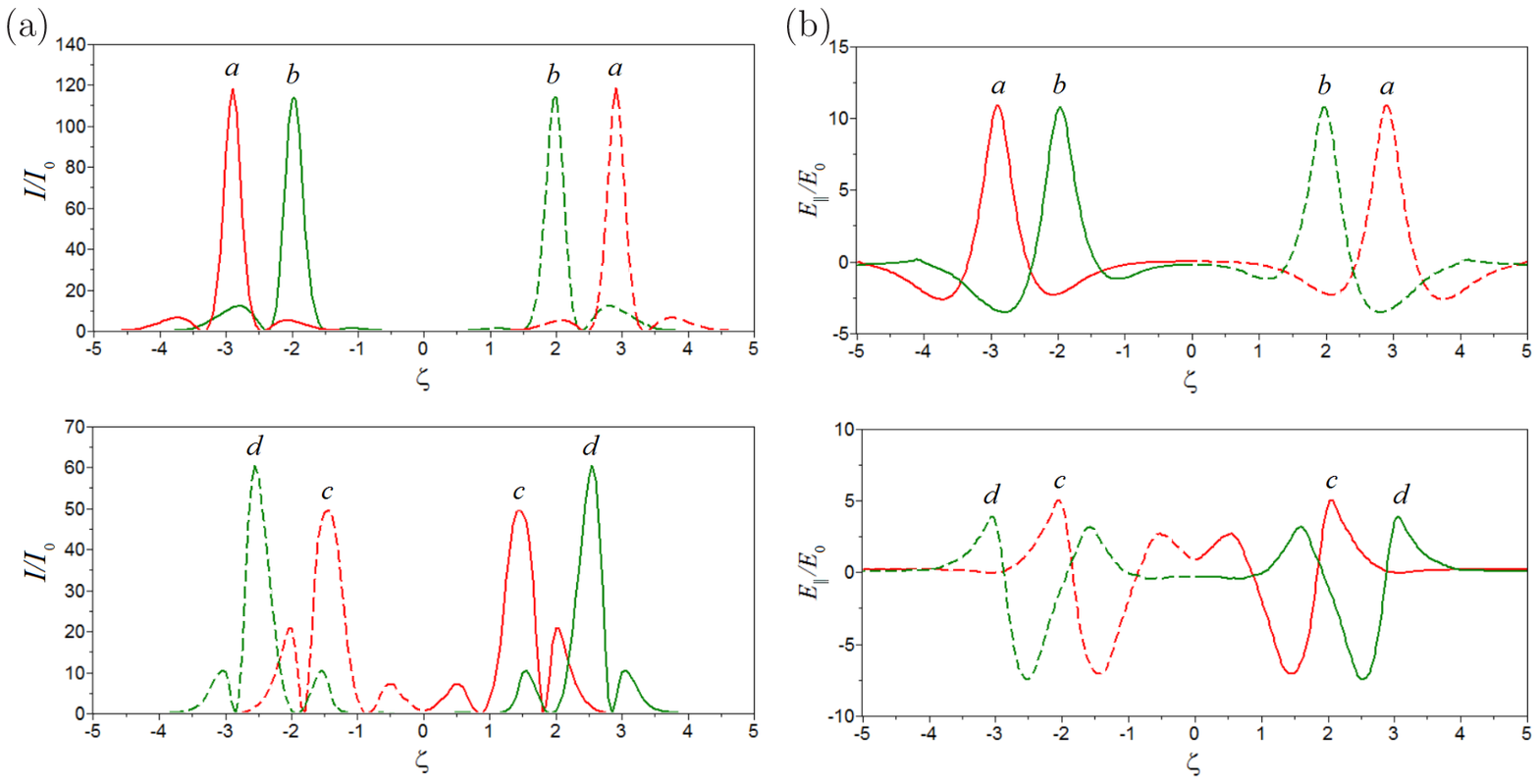}
  \caption{(a): The intensity distribution of the pulses on the $\protect\zeta
    $ axis at different times $\protect\tau $---before the collision, at
    $\protect%
    \tau =0.1$ (line $a$), $\protect\tau =1.0$ (line $b$), and after the
    collision, at $%
    \protect\tau =5.0$ (line $c$), $\protect\tau =6.0$ (line $d$). Dashed
    lines correspond to the counter-propagating pulse. (b): The same for the
    scaled electric field, $%
    E_{\Vert }/E_{0}$.}
  \label{fig3}
\end{figure*}

Figures~\ref{fig1}--\ref{fig4} display results of the 3D simulations of the
interaction of bipolar laser pulses in the CNT array with the parameters
defined in the previous subsection. Figure~\ref{fig1} shows the evolution of
the energy density distribution of the electric field, $I(0,0,\zeta ,\tau )$%
, in the array in the course of the propagation and interaction of the pulses
along the $\zeta $ axis, i.e., in the case of $\xi _{01,2}=\upsilon
_{01,2}=0$. With the value of $\omega _{0}$ chosen above, the length unit on
the $\zeta $ axis corresponds to a distance $c/\omega _{0}\approx 4\times
10^{-4}$ cm in physical units, hence, the coordinate interval shown in the
figure, $\left\vert \zeta \right\vert \leq 5$, corresponds to distance $%
\Delta z\approx 4\times 10^{-3}$ cm. Further, the unit of dimensionless time
$\tau =\omega _{0}t/\sqrt{\varepsilon }$ corresponds to physical time $\sqrt{
  \varepsilon }/\omega _{0}\approx 2.8\times 10^{-14}$ s, and the interval
from $\tau =0$ to $\tau =6$, shown on the figure, corresponds to the physical
time interval $\Delta t\approx 1.7\times 10^{-13}$ s \cite{2}. Thus, the
applicability condition for the collisionless approximation, given by
Eq. (\ref{t}), is valid for these results. The simulations reveal that the
colliding electromagnetic pulses temporarily merge, within time interval $%
2.75<\tau <4.25$, which is accompanied by fluctuations of the field energy
density in a limited region of space. At the post-collision stage, the pulses
separate and continue the propagation in a way similar to that before the
collision. Lastly, it is worth adding that with the increase in the velocity,
the collision between pulses becomes more elastic. The basic reason for this
is that the collision time decreases.

Figure \ref{fig2} shows the distribution of the energy density $I(\xi ,0,\zeta
,\tau )$ of the electric field in the array in the course of the propagation
and interaction of the pulses in the plane of $\left( \xi ,\zeta \right) $ (at
$\upsilon =0$): before the collision [Figs.~\ref{fig2}(a) and~\ref{fig2}(b)],
and after the collision [Figs.~\ref{fig2}(c) and~\ref{fig2}(d)]. Note that the distribution of $I(0,\upsilon ,\tau )$ in the plane
of $\left( \upsilon ,\zeta \right) $ is very similar to what is observed in
the plane of $\left( \xi ,\zeta \right) $. The energy density of the field is
represented by ratio $I/I_{\max }^{\left( \xi ,\zeta \right) }$, different
values of which correspond to a variation of colors (flooded contours) from
violet to red, $I_{\max }^{\left( \xi ,\zeta \right) }$ being the maximum
value of the intensity at given time in the plane of $%
\left( \xi ,\zeta \right) $. This figure shows stable propagation of the
pulses in the array, without any conspicuous spreading.

Figure~\ref{fig3} shows, for clarity, the distribution of the energy density,
$I(0,0,\zeta ,\tau )$, and the electric-field amplitude, $E_{\Vert }(0,0,\zeta
,\tau )/E_{0}$, of the pulses along the $\zeta $ axis at some time $\tau
$. Figures \ref{fig1}--\ref{fig3} clearly corroborate that both pulses
propagate quite stably both before and after the collision. By the stability,
we mean that the pulses pass, with a virtually undistorted shape, the distance
considerably greater than their characteristic sizes along the propagation
direction ($\zeta $).

The simulations demonstrate that, in the course of the propagation of the
pulses, their longitudinal and transverse widths may change, following
decrease of the peak energy density. The reduction in the peak density may be
explained by their dispersive spreading in the propagation direction ($%
\zeta $), as well as by diffractive broadening in the orthogonal directions.
Also, a part of the pulses' energy goes into formation of \textquotedblleft
ripples" or \textquotedblleft tails", i.e., emission of small-amplitude
waves. Nevertheless, in the entire parameter region considered in this work, these effects remain small and do not cause destruction of the
pulses. Note also that, in this paper we consider the conservative model, in
which the total energy remains constant, hence, attenuation of the pulses is
not accounted for by dissipative losses.

Considering various factors that cause the change in the shape of the
colliding pulses, we concluded that the pulses propagating toward each other
induce dynamical spatiotemporal perturbations in the electron density and
scalar potential, $\eta $ and $\Phi $ (similar effects in the spatial domain
have been previously considered in Refs.~\cite{21,EPJD}). Thus, each pulse,
coming into the spatial region already visited by the other one, experiences
an impact from the perturbations left by the second pulse. In other words, the
evolution of the electromagnetic field of the pulse is \textit{indirectly%
} affected by the presence of the counterpropagating one, which leads to the
\emph{indirect} interaction between the pulses through the perturbations
induced by them in the electronic subsystem of the CNT array.

\subsection{Redistribution of the electron density}

\begin{figure*}[tbp]
  \includegraphics[width=.9\textwidth]{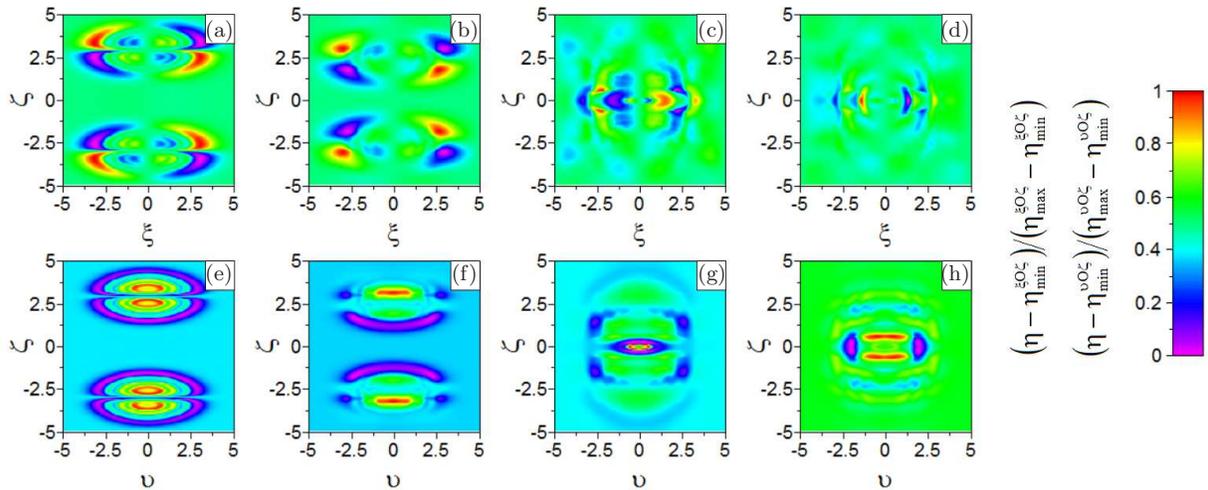}
  \caption{Electron density distribution $\protect\eta (\protect\xi
    ,0,\protect%
    \zeta ,\protect\tau )$ and $\protect\eta (0,\protect\upsilon
    ,\protect\zeta %
    , \protect\tau )$ in planes $\left( \protect\xi ,\protect\zeta \right) $
    and $\left( \protect\upsilon ,\protect\zeta \right) $, respectively,
    before and after the collision of the electromagnetic pulses at different
    times: (a), (e) $%
    \protect\tau =0.1$, (b), (f) $\protect\tau =1.0$, (c) and (g)
    $\protect\tau =5.0$, and (d), (h) $\protect\tau =6.0$. Different values
    of the concentration correspond to a variation of colors from violet
    (minimum) to red (maximum).}
  \label{fig4}
\end{figure*}

As mentioned above, the propagation and interaction of the pulses result in
significant redistribution of the conduction-electrons
concentration. Figure~\ref{fig4} shows the distribution of the electron
density before the collision. The electron density in the planes of $\left(
  \xi ,\zeta \right) $ and $\left( \upsilon ,\zeta \right) $ is represented
by ratios $\left( \eta -\eta _{\min }^{\left( \xi ,\zeta \right) }\right)
/\left( \eta _{\max }^{\left( \xi ,\zeta \right) }-\eta _{\min }^{\left( \xi
      ,\zeta \right) }\right) $ and $\left( \eta -\eta _{\min }^{\left(
      \upsilon ,\zeta \right) }\right) /\left( \eta _{\max }^{\left( \upsilon
      ,\zeta \right) }-_{\min }^{\left( \upsilon ,\zeta \right) }\right) $,
respectively. Different values of the ratios correspond to a variation of
colors from violet to red, similar to Fig.~\ref{fig1}.

\begin{figure*}[tbp]
  \includegraphics[width=.9\textwidth]{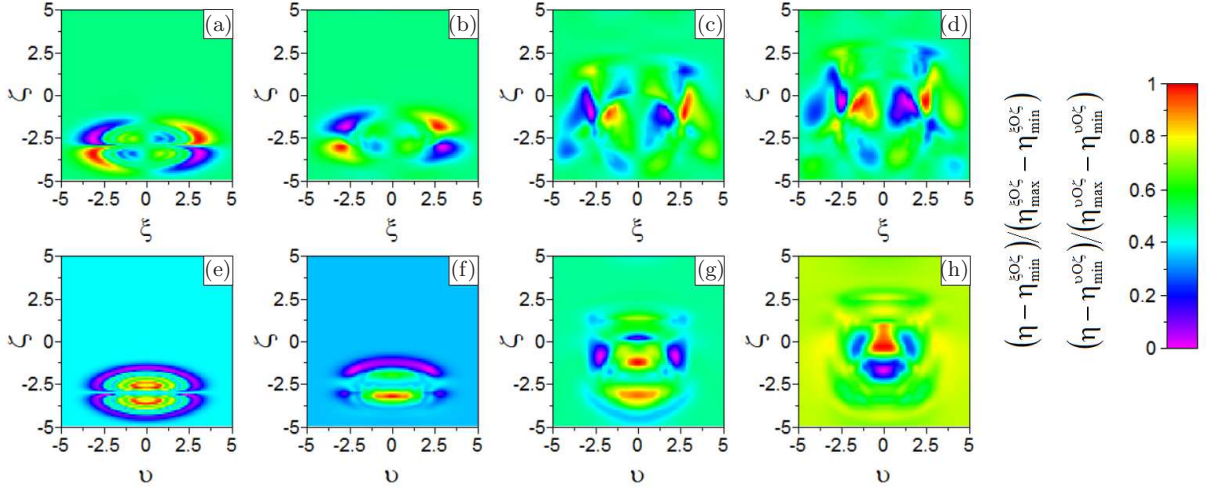}
  \caption{Electron density distributions $\protect\eta (\protect\xi ,0,%
    \protect\zeta ,\protect\tau )$ and $\protect\eta (0,\protect\upsilon ,%
    \protect\zeta ,\protect\tau )$ in the in the planes $\left( \protect\xi ,%
      \protect\zeta \right) $ and $\left( \protect\upsilon ,\protect\zeta
    \right) $%
    , respectively, corresponding to a single propagating pulse, at different
    times: (a), (e) $\protect\tau =0.1$, (b), (f) $\protect\tau =1.0$,
    (c), (g) $\protect\tau %
    =5.0$, and (d), (h) $\protect\tau =6.0$. Different values of the
    density correspond to a variation of colors from violet (minimum) to red
    (maximum).}
  \label{fig5}
\end{figure*}

For comparison, Fig.~\ref{fig5} shows the distribution of the electron density
in the course of the propagation of a single pulse, at the same moments of
time. This figure clearly shows that the propagating pulse leaves behind a
wake consisting of high- and low-density spots. Therefore, during the
collision and thereafter, each pulse not only perturbs the medium, as seen in
Fig. \ref{fig5}, but, as said above, it is forced to propagate in a medium
which has already been perturbed by the counterpropagating pulse, which leads
to the aforementioned indirect interaction between the pulses.  The
electron-concentration distribution after the collision [see Figs.~\ref%
{fig4}(c) and (g) and~\ref{fig4}(d) and (h)] is symmetric with respect
to a plane drawn through the origin perpendicularly to the $\zeta $ axis.

Note that the pulse velocities are high (somewhat smaller than the speed of
light in the surrounding dielectric), therefore, they are actually exposed to
the environment perturbed by the counterpropagating pulse for a very short of
time, the corresponding perturbation in the electron density of the passing
pulse being on the order of a few percent. As a result, the interaction
affects the shape of the pulses, but does not lead to dramatic changes in
their dynamics, and does not destabilize their propagation. Here, we have to
emphasize that the primary goal of this study is to demonstrate that
the pulses survive the collision, which may thus be considered as a
quasi-elastic one. We do not address the stability over much longer
propagation distances, which is a topic for a separate investigation.

\begin{figure}[tbp]
  \includegraphics[width=.48\textwidth]{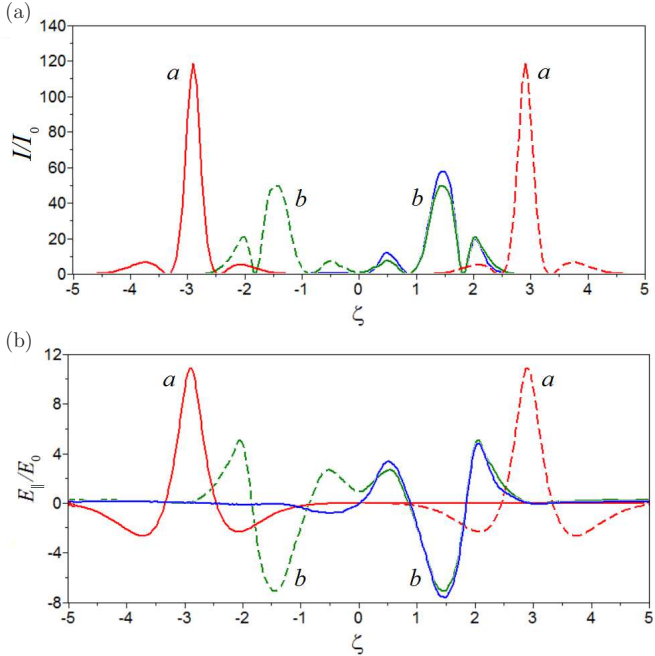}
  \caption{(a): Intensity distribution $I/I_{0}$ for different times. The
    solid red (resp. green) line $a$ (resp. $b$) shows the profile of the pulse
    before (resp. after) the collision at $\protect\tau =0.1$ (resp. at
    $\protect\tau =5.0$). The dashed lines represent the counterpropagating
    pulse at the same instant: red, before collision at $\tau=0.1$, and green,
    after collision at $\tau=5.0$. For the sake of comparison, the profile of
    the single pulse is shown by the blue line at $\protect\tau =5.0$. (b):
    The same for ratio $E_{\Vert }/E_{0}$.}
  \label{fig6}
\end{figure}

Figure \ref{fig6} shows a comparison of quantities $I(0,0,\zeta ,\tau )$ and
$E(0,0,\zeta ,\tau )$ in two situations: (i) the single-pulse propagation, and
(ii) the collision between the pulses. Naturally, the evolution of the shape
of the solitary wave interacting with the other pulse is somewhat different
from the evolution in the case of a single pulse. The differences manifest
themselves in the shape of both the solitary-waves' bodies and their tails
trailing the bodies.

\subsection{Evolution of the Poynting vector}


Finally, we present the results produced by the simulations for the energy
transfer associated with the propagation and interaction of the
electromagnetic pulses in the system. As shown by the simulations, the strong
inequality holds between components of the Poynting vector: $%
|S_{z}|\gg |S_{y}|\gg |S_{x}|$. Thus, the energy transfer is chiefly directed
along the $\zeta $ axis. This result reflects the fact that the processes of
the formation of \textquotedblleft tails" extending in the transverse
direction (along the $\xi $ and $\upsilon $ axes), as well as the transverse
pulse broadening, are much weaker than the longitudinal energy transfer.

Figure \ref{fig7} shows the distribution of the normalized $z$-component of
the Poynting vector, $S_{z}/S_{0}$ (see Appendix~\ref{app:poynting} for
details) at different time instants $\tau $, prior to the collision between
the pulses and afterwards. The figure is an alternative way of presenting the
evolution of the shape of the electromagnetic pulses. For clarity, $%
S_{z}/S_{0}$ is represented by lines of different colors (in the same way as
in Figs.~\ref{fig3} and \ref{fig6}). Specifically, red indicates the profile
of solitary waves before they hit each other, and green refers to the
post-collision stage. Areas on the axis corresponding to the transfer of
energy in the positive and negative directions of the $\zeta $ axis are
indicated by solid and dashed lines, respectively. The figure corroborates
that the pulses retain their individuality after the collision, passing
distances much greater than their own sizes.

Figure \ref{fig7} allows us to evaluate the pulse's energy flux along the
propagation direction. For the given values of parameters, we find [see an
explanation for Eq. \eqref{B5} in Appendix B] $S_{0}\approx 6.5\times 10^{9}$
W/cm$^{2}$, which corresponds to the unit on the vertical axis of Fig. \ref%
{fig7}. Thus, the maxima of $|S_{z}|$ on the $\zeta $ axis for each pulse
before and after the collision yields $7.1\times 10^{11}$ W/cm$^{2}$ and $%
3.2\times 10^{11}$ W/cm$^{2}$, respectively.
\begin{figure}[tbp]
  \includegraphics[width=0.48\textwidth]{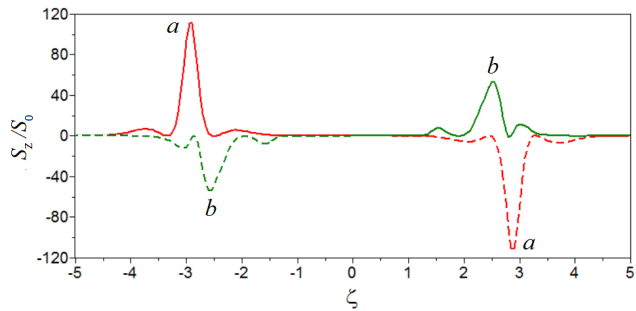}
  \caption{The normalized Poynting-vector component, $S_{z}/S_{0}$, associated
    with the pulses' fields at different times $\protect\tau $. The solid red
    (resp. green) line $a$ (resp. $b$) shows the profile of the pule before
    (resp. after) the collision at $\protect\tau =0.1$ (resp. at $\protect\tau
    =6.0$). The dashed lines represent the counterpropagating pulse at the
    same instant: red, before collision at $\tau=0.1$, and green, after
    collision at $\tau=5.0$. Solid and dashed lines correspond, respectively,
    to the energy transfer in the positive and negative directions along the
    $\protect\zeta $ axis.}
  \label{fig7}
\end{figure}

Some peculiarities of the propagation and interaction of the light bullets
reported in this paper can possibly be used for the design of new elements of
nanoscale optoelectronic devices and laser control systems, as well as for
all-optical data processing. In particular, the effect of the redistribution
of the electron density by an electromagnetic solitary wave suggests
possibilities for the creation of CNT-based light-by-light control devices,
utilizing the respective indirect interaction between the control and signal
pulses. These perspectives for the use of CNT arrays are suggested by results
of Refs.~\cite{41,42} and \cite{Yulin1}-\cite{China}, which predict strong
transformation of radiation as a result of its reflection from solitons in a
nonlinear medium, with dynamical nonuniformity of the refractive index moving
along with the soliton. This effect may be probably used for the design of
metamaterials with rapidly changing dynamical properties.

\section{Conclusions}


Key results of this work may be summarized as follows:

\begin{itemize}
\item[(i)] The complete set
  of equations describing the evolution of the field and charge density were
  derived for the propagation and interaction of light bullets in the array of
  CNTs (semiconducting carbon nanotubes). Our modeling framework takes into
  account the perturbation of the electron density by the nonuniformity of the
  field along the CNT axis.

\item[(ii)] The mechanism of the indirect interaction of extremely short
  electromagnetic pulses via the overlapping perturbations of the electron
  subsystem in the CNT array was thoroughly studied by using the visualization
  of two complementary quantities: the distribution of the energy density of
  the electric field, as a characteristic of the field localization, and
  components of the Poynting vector, as a characteristic of the propagation
  direction and energy flux.

\item[(iii)] The numerical model used in this work allows the investigation of
  different scenarios of the interaction of bipolar electromagnetic pulses in
  the CNT arrays. It has been established that the pulses separate after the
  collision, restoring their shape and steadily moving over distances much
  greater than their characteristic sizes.

\item[(iv)] The electromagnetic pulses induce a dynamic perturbation of the
  electron density in the medium, which, in turn, affects the evolution of the
  electromagnetic waves in the environment. This mechanism may possibly have
  implications for the design of novel optoelectronic devices.
\end{itemize}

Finally, it is worth highlighting that our analysis is limited to the
particular case of a strictly conservative system. Our estimates show that
including effects associated with more realistic composites can lead to slight
quantitative changes in the results, although the qualitative description
remains unchanged.

\section*{Authors' Contribution Statement}
All the authors contributed equally to this work.

\begin{acknowledgments}
  A. V. Zhukov and R. Bouffanais are financially supported by the SUTD-MIT
  International Design Centre (IDC). N. N. Rosanov acknowledges the support
  from the Russian Foundation for Basic Research, Grant No. 16-02-00762, and from
  the Foundation for the Support of Leading Universities of the Russian
  Federation (Grant No. 074-U01). M. B. Belonenko acknowledges support from the
  Russian Foundation for Fundamental Research. E. G. Fedorov is grateful to
  Professor T. Shemesh for his generous support. B. A. Malomed appreciates
  hospitality of the School of Electrical and Electronic Engineering at the
  Nanyang Technological University (Singapore).
\end{acknowledgments}
\pagebreak

\appendix
\begin{widetext}
\section{Derivation of the dispersion relation for linear waves in the system%
}

The full system of evolution equations is
\begin{equation}
  \frac{\partial ^{2}\Psi }{\partial \tau ^{2}}-\left( \frac{\partial ^{2}\Psi
    }{\partial \xi ^{2}}+\frac{\partial ^{2}\Psi }{\partial \upsilon ^{2}}+\frac{
      \partial ^{2}\Psi }{\partial \zeta ^{2}}\right) +\eta \sum_{r=1}^{\infty
  }G_{r}\sin \left[ r\left( \Psi +\int\limits_{0}^{\tau }\frac{\partial \Phi }{
        \partial \xi }d\tau ^{\prime }\right) \right] =0,  \label{AA1}
\end{equation}
\begin{equation}
  \frac{\partial ^{2}\Phi }{\partial \tau ^{2}}-\left( \frac{\partial ^{2}\Phi
    }{\partial \xi ^{2}}+\frac{\partial ^{2}\Phi }{\partial \upsilon ^{2}}+\frac{
      \partial ^{2}\Phi }{\partial \zeta ^{2}}\right) =\beta (\eta -1),
  \label{AA2}
\end{equation}
\begin{equation}
  \frac{\partial \eta }{\partial \tau }=\alpha \sum_{r=1}^{\infty }G_{r}\frac{
    \partial }{\partial \xi }\left\{ \eta \sin \left[ r\left( \Psi
        +\int\limits_{0}^{\tau }\frac{\partial \Phi }{\partial \xi }d\tau ^{\prime
        }\right) \right] \right\} .  \label{AA3}
\end{equation}%
The linearization of Eqs. (\ref{AA1}) and (\ref{AA3}) is performed on top of
the trivial solution, $\Psi =\Phi =0$, $\eta =1$, and (for the time being)
only the first harmonic is kept in the Fourier series in Eqs. (\ref{AA1}) and
(\ref{AA3}) [Eq. (\ref{AA2}) is linear by itself]:%
\begin{equation}
  \frac{\partial ^{2}\Psi }{\partial \tau ^{2}}-\left( \frac{\partial ^{2}\Psi
    }{\partial \xi ^{2}}+\frac{\partial ^{2}\Psi }{\partial \upsilon ^{2}}+\frac{
      \partial ^{2}\Psi }{\partial \zeta ^{2}}\right) +G_{1}\left( \Psi
    +\int\limits_{0}^{\tau }\frac{\partial \Phi }{\partial \xi }d\tau ^{\prime
    }\right) =0,  \label{6lin}
\end{equation}%
\begin{equation}
  \frac{\partial \eta }{\partial \tau }-\alpha G_{1}\frac{\partial }{\partial
    \xi }\left( \Psi +\int\limits_{0}^{\tau }\frac{\partial \Phi }{\partial \xi }
    d\tau ^{\prime }\right) =0.  \label{10lin}
\end{equation}
Further, we define%
\begin{equation}
  \theta \equiv \eta -1,  \label{theta}
\end{equation}
\begin{equation}
  \varphi \equiv \int\limits_{0}^{\tau }\frac{\partial \Phi }{\partial \xi }
  d\tau ^{\prime },~  \label{varphi}
\end{equation}%
hence%
\begin{equation}
  \frac{\partial \Phi }{\partial \xi }\equiv \frac{\partial \varphi }{\partial
    \tau },  \label{Phiphi}
\end{equation}%
and replace Eq. (\ref{AA2}) by the equation differentiated with respect to $%
\xi $:%
\begin{align}
  \beta \frac{\partial \theta }{\partial \xi }-\frac{\partial ^{2}}{\partial
    \tau ^{2}}\left( \frac{\partial \Phi }{\partial \xi }\right) \nonumber\\
+\frac{\partial
    ^{2}}{\partial \xi ^{2}}\left( \frac{\partial \Phi }{\partial \xi }\right) +
  \frac{\partial ^{2}}{\partial \upsilon ^{2}}\left( \frac{\partial \Phi }{
      \partial \xi }\right) +\frac{\partial ^{2}}{\partial \zeta ^{2}}\left( \frac{
      \partial \Phi }{\partial \xi }\right) =0.  \label{8b}
\end{align}%
Then, $\partial \Phi /\partial \xi $ in Eq. (\ref{8b}) is replaced by $%
\partial \varphi /\partial \tau $, according to Eqs. (\ref{Phiphi}), and
definition (\ref{varphi}) is used in Eqs. (\ref{6lin}) and (\ref{10lin}).
Thus, the linearized equations are cast in their final form:%
\begin{equation}
  \frac{\partial \theta }{\partial \tau }-\alpha G_{1}\frac{\partial \varphi }{
    \partial \xi }-\alpha G_{1}\frac{\partial \Psi }{\partial \xi }=0.
  \label{lin1}
\end{equation}%
\begin{equation}
  \beta \frac{\partial \theta }{\partial \xi }-\frac{\partial ^{3}\varphi }{
    \partial \tau ^{3}}+\frac{\partial ^{3}\varphi }{\partial \tau \partial \xi
    ^{2}}+\frac{\partial ^{3}\varphi }{\partial \tau \partial \upsilon ^{2}}+
  \frac{\partial ^{3}\varphi }{\partial \tau \partial \zeta ^{2}}=0,
  \label{lin2}
\end{equation}%
\begin{equation}
  G_{1}\varphi +G_{1}\Psi +\frac{\partial ^{2}\Psi }{\partial \tau ^{2}}
  -\left( \frac{\partial ^{2}\Psi }{\partial \xi ^{2}}+\frac{\partial ^{2}\Psi
    }{\partial \upsilon ^{2}}+\frac{\partial ^{2}\Psi }{\partial \zeta ^{2}}
  \right) =0.  \label{lin3}
\end{equation}

Solutions to linearized equations (\ref{lin1}) and (\ref{lin2}) are looked for in
the usual form,%
\begin{equation}
  \left( \theta ,\varphi ,\Psi \right) \sim \exp \left( -iF\tau +iK\xi
    +iP\upsilon +iQ\zeta \right) ,  \label{exp}
\end{equation}%
where $K,P,Q$ are arbitrary wave numbers, and $F$ is the frequency to be
found. The substitution of ansatz~\eqref{exp} in Eqs.~\eqref{lin1} and~\eqref%
{lin2} leads to the dispersion equation, written in the determinant form:%
\begin{equation}
  \left\vert
    \begin{array}{ccc}
      -iF & -i\alpha G_{1}K & -i\alpha G_{1}K \\
      i\beta K & iF^{3}+iF\left( K^{2}+P^{2}+Q^{2}\right) & 0 \\
      0 & G_{1} & G_{1}-F^{2}+\left( K^{2}+P^{2}+Q^{2}\right)%
    \end{array}%
  \right\vert =0.  \label{det}
\end{equation}%
In an explicit form, Eq. (\ref{det}) is a cubic equation for $F^{2}$:{}$%
\allowbreak $%
\begin{equation}
  \begin{split}
    & \left( F^{2}\right) ^{3}-G_{1}\left( F^{2}\right) ^{2} \\
    & -F^{2}\left[ \left( K^{4}+P^{4}+Q^{4}\right) +2\left(
        K^{2}P^{2}+K^{2}Q^{2}+P^{2}Q^{2}\right) +G_{1}\left(
        K^{2}+P^{2}+Q^{2}\right) +\alpha \beta G_{1}K^{2}\right] \\
    & +\alpha \beta G_{1}\left( K^{4}+K^{2}P^{2}+\allowbreak K^{2}Q^{2}\right)
    =0.
  \end{split}
  \label{cubic}
\end{equation}%
In the limit of $K^{2},P^{2},Q^{2}\rightarrow \infty $, a relevant dispersion
branch is given by an asymptotic solution to Eq. (\ref{cubic})%
\begin{equation}
  F^{2}\approx \sqrt{\left( K^{4}+P^{4}+Q^{4}\right) +2\left(
      K^{2}P^{2}+K^{2}Q^{2}+P^{2}Q^{2}\right) }.  \label{+}
\end{equation}%
%

The main objective is to obtain a band gap from a numerical analysis of Eq. (%
\ref{cubic}), that would be a habitat for solitons. It seems that the band gap
\emph{does not exist}, because, in the limit of $K^{2},P^{2},Q^{2}%
\rightarrow 0$, the asymptotic form of Eq. (\ref{cubic}) is
\begin{equation}
  \left( F^{2}\right) ^{2}+F^{2}\left[ \left( K^{2}+P^{2}+Q^{2}\right) +\alpha
    \beta K^{2}\right] -\alpha \beta K^{2}\left( K^{2}+P^{2}+Q^{2}\right) =0,
  \label{quadr}
\end{equation}%
Obviously, Eq. (\ref{quadr}) shows that, in this limit, the spectrum does not
contain a band gap, but rather a Dirac cone (which is not surprising for a
medium related to graphene). Sometimes, solitons may actually exist or
\textquotedblleft almost exist" in the absence of a true band gap \cite{Nikos}%
. In fact, it is a separate problem to check if solitons exist in the present
model in the strict mathematical sense.

\section{Computation of the conduction current}


Here we aim to derive an expression for the conduction-current density $j$
along the CNT axis, applying an approach similar to the one used in Refs.~%
\cite{34,35} for semiconductor superlattices. Assuming the variation length of
the electromagnetic field to be much larger than the electron de Broglie
wavelength and length $d_{x}$ in the electron dispersion law~\eqref{1}, we
write the current density, associated with the motion of the conduction
electrons, as
\begin{equation}
  j=2e\sum_{s=1}^{m}\int\limits_{-\pi \hbar /d}^{+\pi \hbar
    /d}v_{x}(p_{x}+p_{0},s)f(p_{x},s)dp_{x}.  \label{A1}
\end{equation}%
Here $v_{x}$ is the electron's velocity, and $f(p_{x},s)$ is the associated
distribution function. The integration with respect to quasimomentum $p_{x}$
is carried out over the interval between $-\pi \hbar /d$ and $+\pi \hbar /d$
(note that $p_{x}/\hbar $ thus varies within the first Brillouin zone), and $%
p_{0}$ is determined by the equation of motion,
\begin{equation}
  \frac{dp_{0}}{dt}=-\left(\frac{e}{c}\frac{\partial A}{\partial t}+e\frac{\partial
      \phi }{\partial x}\right).  \label{A2}
\end{equation}%
It follows from Eq.~\eqref{A2} that
\begin{equation}
  p_{0}=-\left(\frac{e}{c}A+e\int_{0}^{t}\frac{\partial \phi }{\partial x}dt^{\prime
    }\right).  \label{A3}
\end{equation}%
Next, to continue the derivation of the current density by means of Eq. %
\eqref{A1}, we need an expression for the velocity of the electrons, $%
v_{x}(p_{x}+p_{0},s)$. To this end, we use the known definition, $%
v_{x}(p_{x},s)=\partial \epsilon (p_{x},s)/\partial p_{x}$. The expression for
$v_{x}(p_{x},s)$ being available, we make a substitution, $%
p_{x}\rightarrow p_{x}+p_{0}$, to obtain $v_{x}(p_{x}+p_{0},s)$.

Next we expand the electron spectrum $\epsilon (p_{x},s)$ in the Fourier
series:
\begin{equation}
  \epsilon (p_{x},s)=\frac{\delta _{0,s}}{2}+\sum_{r=1}^{\infty }\delta
  _{r,s}\cos \left( r\frac{d_{x}}{\hbar }p_{x}\right) ,  \label{A4}
\end{equation}%
\begin{equation}
  \delta _{r,s}=\frac{d_{x}}{\pi \hbar }\int\limits_{-\pi \hbar /d}^{\pi \hbar
    /d}\epsilon (p_{x},s)\cos \left( r\frac{d_{x}}{\hbar }p_{x}\right) dp_{x}~.
  \label{A5}
\end{equation}%
Using expression \eqref{A4} for the electron energy spectrum, we rewrite the
electron velocity $v_{x}(p_{x},s)$ as
\begin{equation}
  v_{x}(p_{x},s)=\frac{\partial \epsilon (p_{x},s)}{\partial p_{x}}=-\frac{%
    d_{x}}{\hbar }\sum_{r=1}^{\infty }r\delta _{r,s}\sin \left( r\frac{d_{x}}{%
      \hbar }p_{x}\right) .  \label{A6}
\end{equation}%
Making the substitution $p_{x}\rightarrow p_{x}+p_{0}$ in Eq.~\eqref{A6}, we
can rewrite current density \eqref{A1}:
\begin{equation}
  j=-2e\frac{d_{x}}{\hbar }\sum_{s=1}^{m}\int\limits_{-\pi \hbar /d}^{\pi \hbar
    /d}\sum_{r=1}^{\infty }r\delta _{r,s}\sin \left\{ r\frac{d_{x}}{\hbar }%
    (p_{x}+p_{0})\right\} f(p_{x},s)dp_{x}.  \label{A7}
\end{equation}%
Further, we make use of the Fermi-Dirac distribution
\begin{equation}
  f(p_{x},s)=\frac{N}{1+\exp \left( \frac{\epsilon (p_{x},s)}{k_{B}T}\right) },
  \label{A8}
\end{equation}%
where $N$ is the constant determined from the normalization condition,
\begin{equation}
  2\sum_{s=1}^{m}\int\limits_{-\pi \hbar /d}^{\pi \hbar /d}f(p_{x},s)dp_{x}=n,
  \label{A9}
\end{equation}%
where pre-factor $2$ accounts for the two possible electron spin
projections. We stress that the conduction-electron density $n$ in Eq.~%
\eqref{A9} is, generally, a function of the spatial coordinates and time, $%
n=n(x,z,t)$.

Now we transform the current density given by Eq.~\eqref{A7}, taking into
regard the distribution function \eqref{A8} and normalization condition %
\eqref{A9}:
\begin{equation}
  j=-\gamma _{0}e\frac{d_{x}}{\hbar }n\sum_{r=1}^{\infty }G_{r}\sin \left\{ r
    \frac{d_{x}}{\hbar }\left( \frac{e}{c}A+e\int\limits_{0}^{t}\frac{\partial
        \phi }{\partial x}dt^{\prime }\right) \right\} ,  \label{A10}
\end{equation}%
\begin{equation}
  G_{r}=-r\frac{\sum_{s=1}^{m}\left( \delta _{r,s}/\gamma _{0}\right)
    \int_{-\pi \hbar /d}^{+\pi \hbar /d}\cos \left( r\frac{d_{x}}{\hbar }
      p_{x}\right) \left[ 1+\exp \left( \frac{\epsilon (p_{x},s)}{k_{B}T}\right) %
    \right] ^{-1}dp_{x}}{\sum_{s=1}^{m}\int_{-\pi \hbar /d}^{+\pi \hbar
      /d}\left\{ 1+\exp \left( \frac{\epsilon (p_{x},s)}{k_{B}T}\right) \right\}
    ^{-1}dp_{x}}.  \label{A11}
\end{equation}%
Using an expression for the electron spectrum following from Eq.~\eqref{A4},
we finally obtain
\begin{equation}
  G_{r}=-r\frac{\sum_{s=1}^{m}\left( \delta _{r,s}/\gamma _{0}\right)
    \int_{-\pi \hbar /d}^{+\pi \hbar /d}\cos \left( r\frac{d_{x}}{\hbar }
      p_{x}\right) \left[ 1+\exp \left( \frac{\delta _{0,s}}{2k_{B}T}
        +\sum_{q=1}^{\infty }\frac{\delta _{q,s}}{k_{B}T}\cos \left( q\frac{d_{x}}{
            \hbar }p_{x}\right) \right) \right] ^{-1}dp_{x}}{\sum_{s=1}^{m}\int_{-\pi
      \hbar /d}^{+\pi \hbar /d}\left[ 1+\exp \left( \frac{\delta _{0,s}}{2k_{B}T}
        +\sum_{q=1}^{\infty }\frac{\delta _{q,s}}{k_{B}T}\cos \left( q\frac{d_{x}}{
            \hbar }p_{x}\right) \right) \right] ^{-1}dp_{x}}.  \label{A12}
\end{equation}%
Finally, with substitution $p_{x}d_{x}/\hbar \rightarrow \kappa $ and notation
$\theta _{r,s}=\delta _{r,s}(k_{B}T)^{-1}$, we obtain expression %
\eqref{4} used in the main text.

\section{The Poynting vector}

\label{app:poynting}

The redistribution of the energy density of energy in the system in the course
of motion and transformation of the solitary waves can be described by
analyzing the evolution of the Poynting vector, i.e., the vector of the
energy-flux density of the electromagnetic field \cite{32,33}:
\begin{equation}
  \mathbf{S}=\left( c/4\pi \right) \mathbf{E}\times \mathbf{H}.  \label{B1}
\end{equation}%
We assume that the medium under consideration is nonmagnetic, with relative
permeability $1$. In this case, the magnetic component of the field, being
expressed in terms of the vector potential as $\mathbf{H}=\nabla \times
\mathbf{A}$, can be written as
\begin{equation}
  \mathbf{H}=H_{0} \left( 0,\frac{\partial \Psi }{\partial \zeta },-%
    \frac{\partial \Psi }{\partial \upsilon }\right) ,  \label{B2}
\end{equation}%
where $H_{0}=-\hbar \omega _{0}(ed_{x})^{-1}$.

Thus, calculating the vector product of the electric \eqref{11} and magnetic %
\eqref{B2} fields, we find the Poynting vector, $\mathbf{S}%
=\{S_{x},S_{y},S_{z}\}$, with the following components:
\begin{align}
  S_{x}& =S_{0} \left( \frac{\partial \Phi }{\partial \upsilon }\frac{
      \partial \Psi }{\partial \upsilon }+\frac{\partial \Phi }{\partial \zeta
    }
    \frac{\partial \Psi }{\partial \zeta }\right) ,  \label{B3} \\
  S_{y}& =-S_{0}\frac{\partial \Psi }{\partial \upsilon } \left( \frac{
      \partial \Psi }{\partial \tau }+\frac{\partial \Phi }{\partial \xi
    }\right) ,
  \label{B4} \\
  S_{z}& =-S_{0}\frac{\partial \Psi }{\partial \zeta } \left( \frac{
      \partial \Psi }{\partial \tau }+\frac{\partial \Phi }{\partial \xi
    }\right) ,
  \label{B5}
\end{align}%
where $S_{0}=\left( c/4\pi \right) (\hbar \omega
_{0})^{2}(ed_{x})^{-2}\varepsilon ^{-1/2}$.

Computation of the absolute value of the Poynting vector allows us to find the
density of the field's energy flux at each point and any instant of time. The
direction of the power transfer is determined by signs of the components of
$\mathbf{S}$: $\text{sign}(S_{x})$, $\text{sign}(S_{y})$, $%
\text{sign}(S_{z})$. The energy transfer associated with the motion of the
electromagnetic pulses and the formation of their tails\ along the $\zeta $
axis is determined by the corresponding component $S_{z}$, given by Eq. %
\eqref{B5}. The energy transfer along the $\zeta $ and $\upsilon $ axes, which
can also contribute to a change in the shape of the pulse, is determined by
components $S_{x}$ [\eqref{B3}] and $S_{y}$ [\eqref{B4}], respectively. The
simulations show that the strong inequality between projections holds,
$|S_{z}|\gg |S_{y}|\gg |S_{x}|$, hence, the energy transfer occurs chiefly
along the $\zeta $ axis. This result reflects the fact that the formation of
the tails in the transverse directions (along $%
\xi $ and $\upsilon $), as well as transverse pulse broadening, are much
weaker than the longitudinal energy transfer associated with the propagation
of the pulses along the $\zeta $ axis.
\end{widetext}


\begin{thebibliography}{99}
\bibitem{1} R. H. Baughman, A. A. Zakhidov, and W. A. de Heer, Science
  \textbf{\ 297}, 787 (2002).

\bibitem{2} S. Iijima, Nature \textbf{354}, 56 (1991).

\bibitem{3} S. Iijima and T. Ichihashi, Nature \textbf{363}, 603 (1993).

\bibitem{4} M. S. Dresselhaus, G. Dresselhaus, P. Eklund, \textit{The science
    of fullerenes and carbon nanotubes} (Elsevier, 1996).

\bibitem{5} \textit{Carbon nanotubes, preparation and properties}, T. W.
  Ebbesen, Ed. (CRC Press, 1996).

\bibitem{6} R. Saito, G. Dresselhaus, and M. S. Dresselhaus, \textit{\
    Physical properties of carbon nanotubes} (World Scientific, 1998).

\bibitem{7} P. J. F. Harris, \textit{Carbon Nanotubes and Related Structures:
    New Materials for the Twenty-First Century} (Cambridge University Press,
  1999).

\bibitem{8} S. A. Maksimenko and G. Ya. Slepyan, J. Comm. Techn. Elect.
  \textbf{\ 47}, 261 (2002).

\bibitem{9} S. A. Maksimenko and G. Ya. Slepyan, in \textit{Handbook of
    Nanotechnology. Nanometer Structure: Theory, Modeling, and Simulation}
  (SPIE Press, Bellingham, 2004).

\bibitem{9A} A. Akhmanov, V. A. Vysloukh, and A. S. Chirkin, \textit{Optics of
    Femtosecond Laser Pulses} (AIP, New York, 1992).

\bibitem{10} F. G. Bass, A. A. Bulgakov, A. P. Tetervov, \textit{\
    High-frequency properties of semiconductors with superlattices} (Nauka,
  Moscow, 1989).

\bibitem{11} M. A. Herman, \textit{Semiconductor Superlattices}
  (Akademie-Verlag, 1986).

\bibitem{12} M. B. Belonenko, S. Yu. Glazov, and N. E. Meshcheryakova, Bull.
  Rus. Acad. Sci: Physics \textbf{73}, 1601 (2009).

\bibitem{13} M. B. Belonenko, N. G. Lebedev, and N. N. Yanyushkina, J. Rus.
  Laser Res. \textbf{31}, 410 (2010).

\bibitem{14} A. V. Zhukov, R. Bouffanais, M. B. Belonenko, and E. G.  Fedorov,
  Mod. Phys. Lett. B \textbf{27}, 1350045 (2013).

\bibitem{15} M. B. Belonenko, E. G. Fedorov, Rus. Phys. J. \textbf{55}, 83
  (2012).

\bibitem{16} M. B. Belonenko, E. V. Demushkina, and N. G. Lebedev, J. Rus.
  Laser Res. \textbf{27}, 457 (2006)

\bibitem{17} A. M. Zheltikov, Phys.-Usp. \textbf{50}, 705 (2007).

\bibitem{R1} S. V. Sazonov, Bull. Rus. Acad. Sci.: Physics \textbf{75}, 157
  (2011).

\bibitem{R2} H. Leblond and D. Mihalache, Phys. Rep. \textbf{523}, 61 (2013).

\bibitem{R3} G. Mourou, S. Mironov, E. Khazanov, and A. Sergeev, Eur. Phys.
  J. Special Topics, \textbf{223}, 1181 (2014).

\bibitem{R4} M. Kolesik and J. V. Moloney, Rep. Prog. Phys. \textbf{77},
  016401 (2014).

\bibitem{R5} D. J. Frantzeskakis, H. Leblond, and D. Mihalache, Rom. J.
  Phys. \textbf{59}, 767 (2014).

\bibitem{18} M. B. Belonenko, N. G. Lebedev, and A. S. Popov, JETP Lett.
  \textbf{\ 91}, 461 (2010).

\bibitem{19} E. G. Fedorov, A. V. Zhukov, M. B. Belonenko, and T. F. George,
  Eur. Phys. J. D \textbf{66}, 219 (2012).

\bibitem{20} H. Leblond and D. Mihalache, Phys. Rev. A \textbf{86}, 043832
  (2012).

\bibitem{LB1} B. A. Malomed, D. Mihalache, F. Wise, and L. Torner, J. Opt.
  B.: Quantum Semiclass. Opt. \textbf{7}, R53 (2005).

\bibitem{LB2} D. Mihalache, Rom. J. Phys. \textbf{57}, 352 (2012).

\bibitem{LB3} D. Mihalache, Rom. J. Phys. \textbf{59}, 295 (2014).

\bibitem{21} A. V. Zhukov, R. Bouffanais, E. G. Fedorov, and M. B.  Belonenko,
  J. Appl. Phys. \textbf{114}, 143106 (2013).

\bibitem{22} M. B. Belonenko, N. G. Lebedev, and N. N. Yanyushkina, Phys.
  Sol. State, \textbf{52}, 1780 (2010).

\bibitem{23} M. B. Belonenko, A. S. Popov, N. G. Lebedev, A. V. Pak, and A.
  V. Zhukov, Phys. Lett. A \textbf{375}, 946 (2011).

\bibitem{24} E. G. Fedorov, N. N. Konobeeva, and M. B. Belonenko, Rus. J.
  Phys. Chem. B, \textbf{8}, 409 (2014).

\bibitem{25} M. B. Belonenko, A. S. Popov, and N. G. Lebedev, Tech. Phys.
  Lett. \textbf{37}, 119 (2011).

\bibitem{26} A. S. Popov, M. B. Belonenko, N. G. Lebedev, A. V. Zhukov, and
  M. Paliy, Eur. Phys. J. D \textbf{65}, 635 (2011).

\bibitem{27} A. S. Popov, M. B. Belonenko, N. G. Lebedev, A. V. Zhukov, and
  T. F. George, Int. J. Theor. Phys. Group Theory Nonlinear Opt. \textbf{15},
  5 (2011).

\bibitem{28} A. V. Zhukov, R. Bouffanais, E. G. Fedorov, and M. B.  Belonenko,
  J. App. Phys. \textbf{115}, 203109 (2014).

\bibitem{EPJD} A. V. Zhukov, R. Bouffanais, H. Leblond, D. Mihalache, E. G.
  Fedorov, and M. B. Belonenko, Eur. Phys. J. D \textbf{69}, 242 (2015).



\bibitem{31} M. B. Belonenko, S. Yu. Glazov, and N. E. Meshcheryakova, Opt.
  Spectr. \textbf{108}, 774 (2010).

\bibitem{32} L. D. Landau and E. M. Lifshitz, \textit{The Classical Theory of
    Fields}, 4th Ed. (Butterworth-Heinemann, Oxford, 2000).

\bibitem{33} L. D. Landau, E. M. Lifshitz, and L. P. Pitaevskii, \textit{\
    Electrodynamics of Continuous Media}, 2nd Ed. (Elsevier, Oxford, 2004).

\bibitem{Vlasov} W. Baumjohann and R. A. Treumann, \textit{Basic Space Plasma
    Physics} (Imperial College Press, London, 1997).

\bibitem{Shanghai} G. Dong, J. Zhu, W. Zhang, and B. A. Malomed, Phys. Rev.
  Lett. \textbf{110}, 250401 (2013); J. Qin, G. Dong, and B. A. Malomed,
  \textit{ibid}. \textbf{115}, 023901 (2015).

\bibitem{34} E. M. Epshtein, Fiz. Tverd. Tela \textbf{19}, 3456 (1976).

\bibitem{35} E. M. Epshtein, Fiz. Tech. Polupr. \textbf{14}, 2422 (1980)
  [Sov. Phys. Semiconductors 14, 1438 (1980)].


\bibitem{37} A. N. Pikhtin, \textit{Optical and Quantum Electronics} (High
  School Publishers, Moscow, 2001).

\bibitem{38} \textit{Femtosecond Laser Pulses: Principles and Experiments},
  Claude Rulli{\'e}re, ed. (Springer-Verlag, Berlin, 1998).

\bibitem{39} Yu. S. Kivshar and B. A. Malomed, Rev. Mod. Phys. \textbf{61},
  763 (1989).

\bibitem{40} J. W. Thomas, \textit{Numerical Partial Differential Equations --
    Finite Difference Methods} (Springer-Verlag, New York, 1995).

\bibitem{41} N. N. Rosanov, JETP Lett. \textbf{88}, 501 (2008).

\bibitem{42} N. N. Rosanov, JETP Lett. \textbf{90}, 428 (2009).

\bibitem{Yulin1} A. Efimov, A. V. Yulin, D. V. Skryabin, J. C. Knight, N.
  Joly, F. G. Omenetto, A. J. Taylor, and P. Russell,
  Phys. Rev. Lett. \textbf{%
    \ 95}, 213902 (2005).

\bibitem{Yulin2} A. Efimov, A. J. Taylor, A. V. Yulin, D. V. Skryabin, and
  J. C. Knight, Opt. Lett. \textbf{31}, 1624 (2006).

\bibitem{Yulin3} R. Driben, A. V. Yulin, A. Efimov, and B. A. Malomed, Opt.
  Exp. \textbf{21}, 19091 (2013).

\bibitem{Yulin4} I. Oreshnikov, R. Driben, and A. V. Yulin, Opt. Lett.
  \textbf{40}, 5554 (2015).

\bibitem{China} Z. Deng, X. Fu, J. Liu, C. Zhao, and S. Wen, Opt. Exp.
  \textbf{24}, 10302 (2016).

\bibitem{Nikos} N. Flytzanis and B.A. Malomed, Phys. Lett. A \textbf{227},
  335-339 (1997).
\end{thebibliography}
\end{document}